\documentstyle[preprint,prb,aps,epsf]{revtex}
\def\rect#1#2{{\vcenter{\vbox{\hrule height.3pt
            \hbox{\vrule width.3pt height#2truecm \kern#1truecm
            \vrule width.3pt}
            \hrule height.3pt}}}}
\def\square{\rect{0.15}{0.15}}

\def\inseps#1#2{\def\epsfsize##1##2{#2##1} \centerline{\epsfbox{#1}}}
\begin{document}
\draft
 
\title{Correlation functions by Cluster Variation Method
for  Ising model with NN, NNN and Plaquette interactions}
\author{E.N.M. Cirillo}
\address{Universit\'e Paris Sud, Batiment 425, 91405 Orsay Cedex, France.
Emilio.Cirillo@math.u-psud.fr}
\author{G. Gonnella, M. Troccoli}
\address{Dipartimento di Fisica dell'Universit\`a di Bari and 
Istituto Nazionale di Fisica Nucleare, Sezione di Bari, via Amendola
173, 70126 Bari, Italy. E\_mail: gonnella@ba.infn.it, troccoli@bari.infn.it}
\vskip -1.truecm
\author{A. Maritan}
\address{International School for Advanced Studies, via Beirut 4,
34014 Trieste, Italy.}
\maketitle
\begin{abstract}
\vskip -1.cm
We consider the procedure for calculating
the pair correlation function in the context of the
Cluster Variation Methods. As specific cases, we study
the pair correlation function in the paramagnetic phase of the
Ising model with nearest neighbors, next to the nearest
neighbors and plaquette interactions in two and three dimensions.
In presence of competing interactions, the so called disorder line
separates in the paramagnetic phase a region where the correlation function
has the usual exponential behavior from a region where the correlation
has an oscillating exponentially damped behavior.
In two dimensions, using the plaquette as the maximal cluster
of  the CVM approximation,
we calculate the phase diagram and the disorder line for a case where
a comparison is possible with results known in literature
for the eight-vertex model.
In three dimensions, in the CVM cube approximation, we calculate the phase
diagram and the disorder line in some cases of particular interest.
The relevance of our results for experimental systems like mixtures of
oil, water and surfactant is also discussed.
 
{\it Key words: Ising models, CVM, correlation functions, disorder line}
\end{abstract}
\vskip 0.8 cm
\pacs{PACS numbers: 05.50.+q; 68.35.Rh; 75.10.Hk}
 
\section{Introduction}
Statistical systems are exactly solvable only in the simplest cases
and often in unphysical spatial dimensions.
In  three dimensions, in particular, solutions of common statistical models
like the Ising model are not known.
On the other hand, numerical simulations sometimes
are not a very efficient tool for studying systems
with complex phase diagrams where an analytical
control of the phase behavior of the system
would be very helpful.
If one is not interested in the critical behavior
of the system,  a very useful generalization
of  mean field methods for studying the phase diagram of
spin systems is given by the cluster variation method (CVM) \cite{kik1,An,Pel1}.
This method generally gives a very accurate description of the phase 
diagram that can be appreciated 
when a comparison is possible with exact or Monte Carlo
results \cite{Pel2}. It has been extensively used for the calculation 
of equilibrium thermodynamics properties of spin models.

The CVM is based on the application of the variational 
principle of the statistical mechanics with the 
simplification that the density matrix of the system
is factorized over clusters inside which the system is treated
exactly. The size of these cluster, sometimes called 
``maximal clusters'', depends on the accuracy requested and on the necessity
of including all the interaction of the system and of reproducing
the structure of the ground-states. 
Correlation functions have been calculated in the framework of 
CVM approximation only in few simple cases.
Some properties of the pair correlation function
of the  two-dimensional Ising model  
 have been  studied  in the pair (Bethe) and plaquette
 approximation levels of the CVM \cite{san1}.
The CVM has been
also applied to the study of the correlation
function in the disordered phase of the two-dimensional ANNNI
(Axial Next Nearest Neighbors Ising) model \cite{Fin}.

In this paper we present a  procedure 
for the calculation of correlation functions 
that can be generally applied to any 
choice of maximal cluster in the CVM scheme.
It can be also implemented in numerical calculations 
together with the Natural Iteration Method \cite{kik2}
which is a numerical minimization procedure very convenient
for calculating the maximal cluster density matrix  in the most
complex cases. In particular
we apply our procedure to study the behavior 
of the pair correlation functions in the paramagnetic
phase of the  Ising model with nearest neighbors (NN),
next-to-the-nearest neighbors (NNN), and plaquette (P) interactions
in two and three dimensions. In three dimensions, the
phase diagram of this model can be correctly studied 
by implementing the CVM cube approximation with
8 independent site magnetizations in order to take into account the complex
ground-state structure. Our goal will be to calculate the pair correlation 
function in the same approximation used for the phase diagram.

The model with NN, NNN and P interactions  has the same hamiltonian
of the symmetric
two-dimensional eight-vertex model \cite{bax}.  
In three-dimensions this  model is interesting also because
it constitutes
 a  realization of interacting
random surfaces  on a lattice \cite{Kar,m1}.
Indeed, the Peierls interfaces between domains of different sign
can be interpreted as an ensemble of interacting surfaces with a Boltzmann
weight depending on the area, the mean curvature and the length of the
intersection lines between the surfaces \cite{m1}.
It has been shown \cite{m2,cgm} that the phase diagram of this model,
studied in mean field approximation, can well describe the
phase diagram of
experimental systems like fluid mixtures of water, oil and surfactant
or other complex fluids.
Recently, the model with NN, NNN and P interactions has been also put
in relation with string theory
and in particular
with a discretized string model (the so-called gonihedric model) characterized
by a zero energy cost for the area of surfaces \cite{gon}.
The CVM cube approximation has been applied to the study
of the phase diagram of the model with NN, NNN and P interactions
in the  parameter region corresponding to
the gonihedric model \cite{beppe1}.

A question of particular interest in the phase diagram 
of  models with competing interactions is the behavior
 of the pair correlation function in the paramagnetic
phase. Here, generally, the so called ``disorder line''
separates a region where the pair 
correlation 
has the usual exponentially damped behavior from a region
where the correlation develops an oscillatory exponentially
damped behavior \cite{FW}. In the case of the three-dimensional 
model with NN, NNN, P interactions,
this latter region can be identified with 
the microemulsion phase of surfactant systems which is 
a disordered phase but with an ordered structure on short length-scales 
\cite{GS}.
Therefore, in this model, 
the possibility of calculating the correlation functions 
in the same CVM approximation used for the phase diagram is interesting 
also for some physical applications.

The plan of the paper is the following. In Sect. II we briefly summarize
the CVM procedure and discuss how correlation functions
can be calculated. In Sect. III we consider the bidimensional  eight-vertex
model. We calculate the disorder line and
compare our results with other results existing in literature.
In Sect. IV we will carry out the calculation of the correlation functions
in the three-dimensional  case and study the disorder line in two 
cases of particular
physical interest. Some conclusions will follow.

\section{CVM and Correlation Functions}
 
In this section  we first briefly describe the CVM approximation.
We refer to the papers of refs. \cite{An} for further details.
Then we will discuss how the pair correlation 
functions can be evaluated starting from the CVM free energy.

Suppose to have a spin hamiltonian defined on a given lattice
as
\begin{equation}
- \beta{\cal H}(\sigma)=\sum_{\alpha \in {\cal I}} J_{\alpha} \sigma_{\alpha}
\label{eq:genh}
\end{equation}
where $\beta$ is the inverse of the temperature,
$J_{\alpha}$ is the interaction parameter for the cluster 
$\alpha$ and $\sigma_{\alpha}$
 is the product of spins $\sigma_i$ on the sites $ i \in  \alpha$.
The first step of the CVM procedure is to choose
a set of maximal clusters ${\cal M}$ in which the system
is treated exactly such that 
each cluster $\alpha \in {\cal I} $ is 
included in one maximal cluster $M \in  {\cal M}$.
Then the CVM  approximated  free energy 
of the system can be expressed \cite{An}
as the minimum of 
\begin{equation}
\beta {\cal F}=-\sum_{\alpha \in {\cal I}} J_{\alpha}\xi_{\alpha} 
+\sum_{\alpha \subseteq M \in {\cal M}} a_{\alpha} {\rm Tr} \rho_{\alpha} 
\ln \rho_{\alpha}
\label{eq:genf}
\end{equation}
Here the first sum is 
the internal energy of the system with the  $\xi_{\alpha}$ 
being  variational parameters
representing the expectation value of the multisite product of spins 
$\sigma_{\alpha}$.
The second term in the r.h.s of (\ref{eq:genf})
is the entropy of the system and is a sum over
all the clusters which are  possible subclusters of the maximal
clusters.
The set of coefficients $a_{\alpha}$
depends on the lattice structure and on the choice of the maximal cluster;
they can be easily found applying the eqs.(16')  in the paper by An \cite{An}.
The trace Tr is the sum over the allowed configurations and
$\rho_{\alpha}$
is  the density matrix for the cluster $\alpha$.
In the case of  Ising spins $\sigma_i=\pm1$ one can write
\begin{equation}
\rho_{\alpha} = 2^{- n_{\alpha}} [1 + \sum_{\beta \subseteq \alpha}
\sigma_{\beta}\xi_{\beta}]
\label{eq:rh}
\end{equation}
where $n_{\alpha}$ is the number of the sites in  the cluster $\alpha$, 
the sum extends over all the subclusters $\beta$ of the cluster
$\alpha$, and $\sigma_{\beta} = \prod_{i \in \beta} \sigma_i$.
A consequence of eq. (\ref{eq:rh}) is that
 $\xi_{\beta} = {\rm Tr} \sigma_{\beta} \rho_{\beta}$ for a cluster
$\beta \subseteq \alpha$.
The parameters $\xi_{\alpha}$ have to verify the  minimization conditions
\begin{equation}
0 = \frac {\partial {\cal F} } {\partial \xi_{\alpha}}
\label{eq:equih}
\end{equation}

In order to study the pair correlation function for a group of spins
 $\sigma_{\gamma}$, it is convenient
to introduce a local external field in the starting hamiltonian
coupled to all
 clusters of the same type of $\gamma$.
For one of these  clusters the equilibrium condition becomes
\begin{equation}
h_{\gamma} =\frac {\partial {\cal F} }{\partial \xi_{\gamma}}
\label{eq:equihgamma}
\end{equation}
Therefore  the connected
pair correlation function is given by 
\begin{equation} 
<\sigma_{\gamma_0} \sigma_{\gamma{\vec r}}>_c = 
\frac {\partial^2 {\cal F} }
{\partial h_{\gamma_0} \partial h_{\gamma_{\vec r}}} = 
\frac {\partial \xi_{\gamma_0}} { \partial h_{\gamma_{\vec r}}} 
\label{eq:equicorr}
\end{equation}
where we have used  the linear response theorem
and the equilibrium conditions (\ref{eq:equih}).
Now, the free energy ${\cal F}(J_{\alpha},h_{\gamma})$
can be considered formally as a function of the independent
variables $\xi_{\gamma}$ instead of the $h_{\gamma}$.
Therefore  the 
last term of the above equation can 
be evaluated through its  inverse matrix
that  can be calculated 
by differentiating (\ref{eq:equihgamma}) with respect to 
$\xi_{\gamma_{\vec r}}$:
\begin{equation}
\frac {\partial h_{\gamma_0}} {\partial \xi_{\gamma_{\vec r}}} = \frac
{\partial^2  {\cal F}} { \partial \xi_{\gamma_{\vec 0}} 
\partial \xi_{\gamma_{\vec r}}} \equiv {\cal A}_{\gamma_0,\gamma_{\vec r}}
\label{eq:ema}
\end{equation}

Once that the above matrix is obtained,  its Fourier transform
can be simply inverted and will give the requested 
correlation function in Fourier space.
However, except that in the simplest cases,
writing  the matrix ${\cal A}_{\gamma_0,\gamma_{\vec r}}$
can be a complicate task from a computational point of view and
requires some further considerations that we present 
here in a general way and that will become more transparent 
when applied to specific examples in the next sections.
 
In the matrix ${\cal A}_{\gamma_0,\gamma_{\vec r}}$
there appear generally derivatives like
$\partial \xi_{\alpha}/\partial \xi_{\gamma}$ (see, e.g., 
eqs. (\ref{eq:gamma-bruta},\ref{eq:kappa_i}))
where $ \gamma$ is a cluster of different type of $\alpha$.
The evaluation of these derivatives can be performed 
by differentiating
the state equations for  $\xi_{\alpha}$  
with respect to $\xi_{\gamma}$.
This procedure will give a system of linear equations 
\begin{equation}
\frac {\partial^2 {\cal F}} {\partial \xi_{\alpha} \partial \xi_{\gamma}} = 0
\label{eq:caz}
\end{equation}
in the variables $\partial \xi_{\alpha}/\partial \xi_{\gamma}$.
Some of the equations (\ref{eq:caz}) are  homogeneous.
More specifically these  situations may occur:
i) the cluster $\alpha$ belongs only to one maximal cluster.
  In such a case the equation (\ref{eq:caz})
  is  homogeneous or not,
  depending if $\gamma$ is external or not to that maximal cluster.
ii) The cluster $\alpha$ belongs to more than one maximal cluster
   and $\gamma$ is external to the maximal clusters with $\alpha$ in common.
   In such a case the equation  (\ref{eq:caz}) is homogeneous.
iii) The cluster $\alpha$ belongs  to more than one maximal cluster
 and $\gamma$ belongs to one of the maximal clusters with $\alpha$ in common.
  In this  case the equation is generally not homogeneous.
It can be shown  that a solution of the  system (\ref{eq:caz}) 
can be obtained by setting  equal to zero all the 
terms $\partial \xi_{\alpha}/\partial \xi_{\gamma}$ 
appearing in homogeneous equations.
This procedure will be more clear and it will be shown in explicit cases
in the next two sections.

Finally we observe that a great simplification occurs when we want to 
calculate the correlation function in the paramagnetic phase
of a system with only even interactions. In this case
only the derivatives 
$\partial \xi_{\alpha}/\partial \xi_{\gamma}$, 
where $\alpha$ and $\gamma$ have both 
an odd (or even) number of sites,
are different from zero.

\section{Pair Correlation in the 8-vertex model}

In the following of the paper our principal aim will be to calculate
the two-spin 
correlation function and the disorder line
 of the model defined by the hamiltonian
\begin{equation}
-\beta {\cal H} = J_1\sum_{<x,y>}\sigma_x\sigma_y+J_2\sum_{<<x,y>>}
\sigma_x\sigma_y+J_3\sum_{^x_y\square^w_z}\sigma_x\sigma_w\sigma_z\sigma_y
\label{eq:ham-8vert}
\end{equation}
where $\beta$ is the inverse of the temperature and the
$\sigma_x$ are Ising variables defined on the sites of a cubic lattice.
The three sums respectively refer to nearest, next-to-the nearest neighbors
and plaquettes of the lattice.
In this section we  consider the CVM plaquette approximation
of this hamiltonian on  a square lattice
and we use it to calculate 
the phase diagram, 
the correlation function
in the disordered phase and  the disorder line in the case $J_3=0$.
In the following section we will consider the three-dimensional case.
\par
\vskip 1cm
{\it 3.1 The plaquette approximation and the phase diagram.}
\par
The natural choice for studying the phase diagram of the model 
(\ref{eq:ham-8vert}) in two dimensions is 
the CVM approximation level where  
the maximal cluster with independent density matrix
is the plaquette cell of the square lattice.
When  a magnetic-field term  is added to (\ref{eq:ham-8vert}),
the  CVM free-energy density functional ${\cal F}$
to be minimized is given by
\begin{equation}
\begin{array}{ll}
\beta{\cal F} \ &=\ -\; J_1\sum_{<xy>}\, 
{\rm Tr}(\sigma_x\sigma_y\,\rho_{<xy>})\; -\;
J_2\sum_{<<xy>>}\,{\rm Tr}(\sigma_x\sigma_y\,\rho_{<<xy>>})\; + \\
&  -\;J_3\sum_{^x_y\square^w_z}\, {\rm Tr}(\sigma_x\sigma_y\sigma_w\sigma_z
\,\rho_{^x_y\square^w_z})\; 
-\; \beta\sum_x\, h_x{\rm Tr}(\sigma_x\rho_x) \; +\; 
\sum_{^x_y\square^w_z}\,{\rm Tr} {\cal L}(\rho_{^x_y\square^w_z})
+\\
&  -\;\sum_{<xy>}\,{\rm Tr}{\cal L}(\rho_{<xy>})
+\; \sum_x\,{\rm Tr}{\cal L}(\rho_{x})\\
\label{eq:fplac-acca}
\end{array}
\end{equation}
where the spins $\sigma_x$ in 
the argument of the traces are the spins in the cluster one is
considering, ${\cal L}(a)=a\log a$ for a real number $a$ and
 $\rho_{\alpha}$ is the density matrix
related to the cluster of a given type $\alpha$ of the lattice;
 in the following
we need also the matrix $\rho_{[xyz]}$ where $[xyz]$ denotes a three site 
corner cluster.
The density matrix $\rho_{\square}$ is subjected to the constraint
${\rm Tr} \rho_{\square} =1$ and the smaller cluster density matrices
are obtained by partial traces of $\rho_{\square}$.
Notice that we have not
assumed any {\it a priori} symmetry property for our density matrices;
this implies that all possible
states with different magnetic order in a single plaquette 
can be studied by this approximation.
\par
\begin{figure}
\vskip -10mm
\inseps{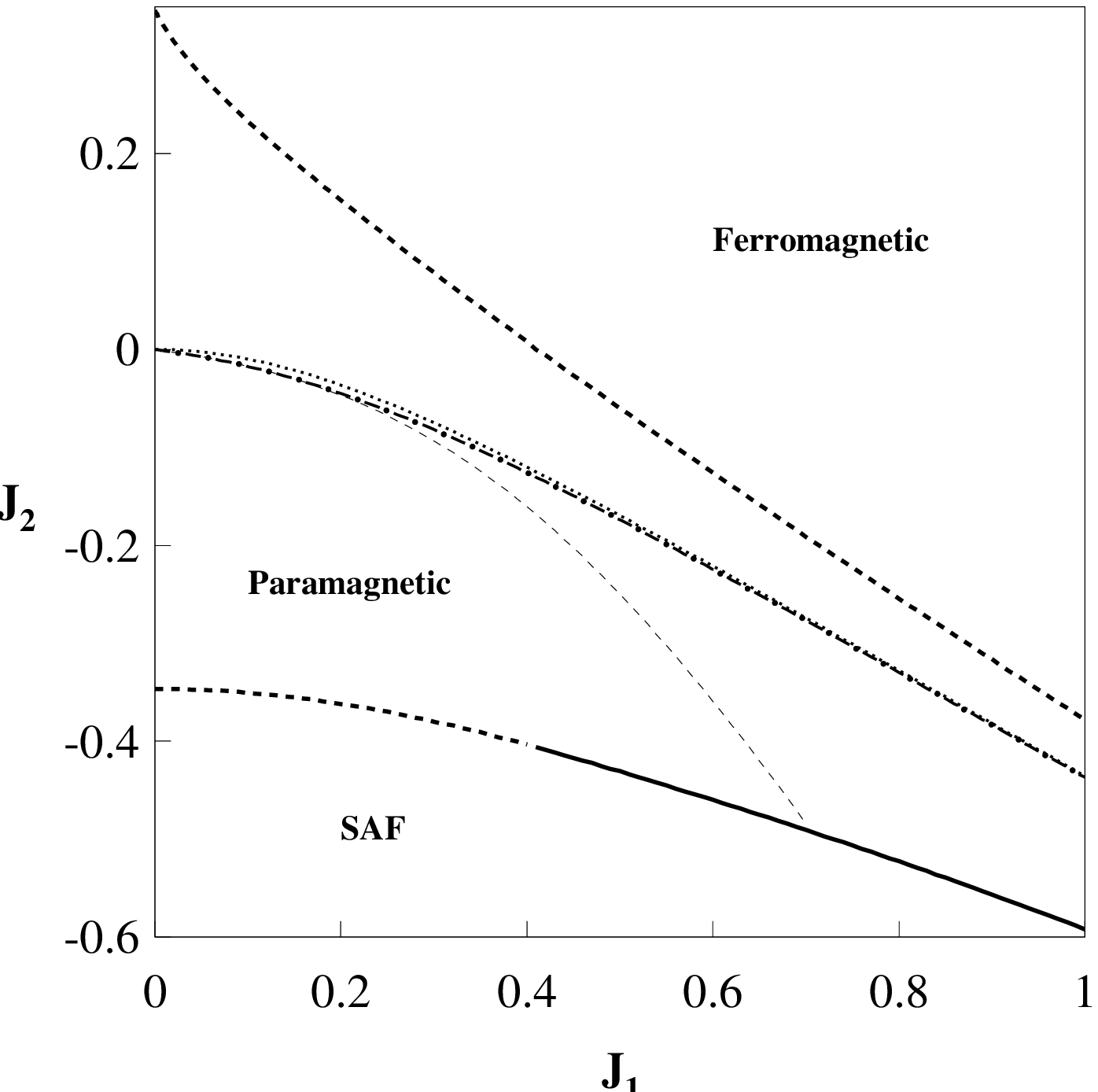}{0.95}
\vskip +20mm
\caption[Testo che va nell'elenco]{
}
\label{fig:diagr-2d}
\end{figure}

The plaquette density matrix $\rho_{\square}$ can be calculated by minimizing
the free energy (\ref{eq:fplac-acca}) via the natural iteration equations
\cite{kik2}; then, after partial traces,
one can work out the phase diagram of the system.
In Fig. \ref{fig:diagr-2d} the portion of the phase diagram with $J_3=0$ is 
shown.
The line separating the ferromagnetic and the paramagnetic phases is
critical, while on the line separating the paramagnetic and the
lamellar phase - sometimes called Super AntiFerromagnetic (SAF) phase -
 there is a tricritical point at $J_1=0.405\pm 0.005, J_2=-0.4045\pm 0.0015$. 
The existence of
a first order phase transition between the paramagnetic
and the lamellar phase is also confirmed by other authors 
\cite{Moran1,Buzzi}  that 
have used the plaquette CVM approximation to study the phase diagram 
of the 8-vertex model.  Our results for the location of the tricritical 
point are in agreement with results of refs. \cite{Moran1,Buzzi}.
The discrepancy of this result with the results of 
 Monte Carlo simulations and  perturbations methods 
where a nonuniversal critical behavior is found 
- for a list of 
references see \cite{Buzzi}, 
may be an artifact of the CVM approximation.
\par  
\vskip 1cm
{\it 3.2 The Correlation Function.}
\par
For the calculation of the correlation function it is convenient to write
the free energy (\ref{eq:fplac-acca}) as a function of the parameters
\begin{equation}
\begin{array}{ll}
m_x&={\rm Tr}(\sigma_x\rho_x)\\
l_{xy}&={\rm Tr}(\sigma_x\sigma_y\rho_{<xy>})\\
c_{xy}&={\rm Tr}(\sigma_x\sigma_y\rho_{<<xy>>})\\
k_{xyz}&={\rm Tr}(\sigma_x\sigma_y\sigma_z\rho_{[xyz]})\\
d_{xywz}&={\rm Tr}(\sigma_x\sigma_y\sigma_w\sigma_z\rho_{^x_y\square^w_z})\\
\end{array}
\label{eq:magn}
\end{equation}
that are related to the density matrices by:
\begin{equation}
\begin{array}{ll}
\rho_x= &\frac{1}{2}(1+m_x\sigma_x)\\
\rho_{<xy>}= &\frac{1}{4}(1+m_x\sigma_x+m_y\sigma_y+l_{xy}\sigma_x\sigma_y)\\
\rho_{<<xy>>}= &\frac{1}{4}(1+m_x\sigma_x+m_y\sigma_y+c_{xy}\sigma_x\sigma_y)\\
\rho_{^x_y\square^w_z}= &\frac{1}{16}(
1+m_x\sigma_x+m_y\sigma_y+m_w\sigma_w+m_z\sigma_z+\\
& l_{xy}\sigma_x\sigma_w+l_{wz}\sigma_w\sigma_z+l_{zy}\sigma_z\sigma_y+
  l_{wx}\sigma_w\sigma_x+\\
& c_{xz}\sigma_x\sigma_z+c_{yw}\sigma_y\sigma_w+\\
& k_{yxw}\sigma_y\sigma_x\sigma_w+
  k_{xwz}\sigma_x\sigma_w\sigma_z+
  k_{wzy}\sigma_w\sigma_z\sigma_y+
  k_{zyx}\sigma_z\sigma_y\sigma_x+\\
& d_{xywz}\sigma_x\sigma_w\sigma_z\sigma_y)\\
\end{array}
\label{eq:matr}
\end{equation}
\par
We have seen in Sect. II that in order to calculate
the pair correlation function, we need 
to evaluate the matrix
\begin{equation}
\left(
\frac{\partial h_x}{\partial m_y}
\right)_{m_z=0\;\forall z}
\label{eq:matrix-acca}
\end{equation}
Therefore we consider the state equation for the magnetization
$\partial {\cal F}/\partial m_i = 0$:
\begin{equation}
\beta h_i=
-\frac{1}{4}\sum_{<iy>}{\rm Tr}(\sigma_i\log\rho_{<iy>})
+\frac{1}{16}\sum_{^i_y\square^w_z}{\rm Tr}(\sigma_i
              \log\rho_{^i_y\square^w_z})
+\frac{1}{2}{\rm Tr}(\sigma_i\log\rho_i)
\label{eq:acca_i}
\end{equation}
where the first and the second sums are respectively taken over the
pairs of nearest neighbors and the plaquettes containing the site $i$.
We start from the diagonal elements  of the matrix 
(\ref{eq:matrix-acca}) 
given by 
\begin{displaymath} 
\beta\frac{\partial h_i}{\partial m_i}=
-\frac{1}{16}\sum_{<iy>}{\rm Tr}\left(\frac{1}{\rho_{<iy>}}\right)
+\frac{1}{4}{\rm Tr}\left(\frac{1}{\rho_i}\right)+
\end{displaymath}
\begin{equation}
+\frac{1}{256}\sum_{^i_y\square^w_z}
{\rm Tr}\left[\frac{\sigma_i}{\rho_{^i_y\square^w_z}}\left(
\sigma_i
+\sigma_y\sigma_z\sigma_w\frac{\partial k_{yzw}}{\partial m_i}
+2\sigma_i\sigma_y\sigma_z\frac{\partial k_{iyz}}{\partial m_i}
+\sigma_w\sigma_i\sigma_y\frac{\partial k_{wiy}}{\partial m_i}
\right)\right] 
\label{eq:gamma-bruta}
\end{equation}
where all the derivatives are evaluated in $m_z=0\;\forall z$
and $\frac{\partial k_{zwi}}{\partial m_i}=
\frac{\partial k_{iwz}}{\partial m_i}$ has been assumed.
Derivatives like $\frac{\partial l_{xy}}{\partial m_i},
\frac{\partial c_{xy}}{\partial m_i},
\frac{\partial d_{xywz}}{\partial m_i}$ do not appear 
because they are odd functions of $m_i$ evaluated at $m_i=0$.
Moreover in the paramagnetic phase 
$l_{xy}=l$, $c_{xy}=c$ and $d_{xywz}=d\;\forall x,y,w,z$.
To simplify the notation we introduce 
\begin{equation}
k_1 =\frac {\partial k_{iyz}}{\partial m_i}\;,\;\; 
k_2 =\frac {\partial k_{wiy}}{\partial m_i}\;\;{\rm and}\;\; 
k_3 =\frac {\partial k_{yzw}}{\partial m_i} 
\label{eq:kappa_i}
\end{equation}

While the parameters $l$, $c$ and $d$ can be  calculated as usual 
by solving the
equilibrium state equations by means of the Natural Iteration Method,
the calculation of $k_1$, $k_2$ and $k_3$ is a little bit more
intriguing. As explained in Sect. II it is  convenient 
to consider the state equation obtained
by setting equal to zero the derivative
of the free energy with respect to $k_{xyz}$ 
\begin{equation}
{\rm Tr}(\sigma_x\sigma_y\sigma_z\log\rho_{^x_y\square^w_z})=0
\label{eq:state-kappa}
\end{equation}
Now, by differentiating this equation with respect to $m_j$ with 
$j=x,y,w$,  we obtain the system:
\begin{equation}
{\rm Tr}\left[\frac{\sigma_x\sigma_y\sigma_z}{\rho_{^x_y\square^w_z}}
\left(\sigma_j+
\sum_{[x',y',z']\subset ^x_y\square^w_z}
\sigma_{x'}\sigma_{y'}\sigma_{z'}
\frac{\partial k_{x'y'z'}}{\partial m_j}
\right)\right]=0\;\;\;\;\; j=x,y,w  
\label{eq:system-kappa}
\end{equation}
By solving this system of three equations in the variables
$k_1$, $k_2$ and $k_3$ one obtains    
\begin{eqnarray}
\lefteqn
k_{\ \,1}\ =\ \frac{A(C-D)}{(D+B+2A)(D+B-2A)}\ \ ;\ \ 
k_2\ =\ \frac{(C-D)(B^2+BD-2A^2)}{(D-B)(D+B+2A)(D+B-2A)}\ \ \nonumber\\
& & \nonumber\\
k_3\ =\ \frac{2A^2(C+D-2B)+(B^2-DC)(B+D)}{(D-B)(D+B+2A)(D+B-2A)}\ \ 
\ \ \ \ \ \ \ \ \ \ \ \ \ \ \ \ \ \ \ 
\label{eq:eqarr0}
\end{eqnarray}
where
\begin{eqnarray}
A\ =\ {\rm Tr}\left(\frac{\sigma_x\sigma_y}{\rho_{^x_y\square^w_z}}
\right)\ \ &;&\ \ 
B\ =\ {\rm Tr}\left(\frac{\sigma_y\sigma_w}{\rho_{^x_y\square^w_z}}
\right)\ \ \nonumber\\
& & \nonumber\\
C\ =\ {\rm Tr}\left(\frac{\sigma_x\sigma_y\sigma_w\sigma_z}
{\rho_{^x_y\square^w_z}}\right)\ \ &;&\ \ 
D\ =\ {\rm Tr}\left(\frac{1}{\rho_{^x_y\square^w_z}}\right)\ \
\label{eq:eqarr1} 
\end{eqnarray}
\par
The insertion of eqs. (\ref{eq:eqarr0},\ref{eq:eqarr1}) 
in (\ref{eq:gamma-bruta})
gives the diagonal elements of (\ref{eq:matrix-acca}). 
Going further, by differentiating (\ref{eq:acca_i})
 with respect to $m_j$ with
$i,j$ nearest neighbours, one realizes that derivatives like 
$\frac{\partial k_{ixy}}{\partial m_j}$ with $j \notin  {^x_y\square^i_z} $
appear.
However, these derivatives are zero.
Indeed, by differentiating the state equation for $k_{xyz}$ with respect
to $m_j$ with $j$ external to the plaquette of $[xyz]$, we obtain an 
 equation similar to (\ref{eq:system-kappa}) but without the first
term $\sigma_j$ in  circular brackets. 
Changing the site $j$ we again obtain a systems 
 for the derivatives $\frac{\partial k_{xyz}}{\partial m_j}$ with $j$
external to the plaquette of $[xyz]$, but in this case these derivatives are 
zero since the system is homogeneous.

A consequence of the above considerations is that 
the matrix (\ref{eq:matrix-acca}) can be written as
\begin{equation}
\beta\left(\frac{\partial h_i}{\partial m_y}\right)_{m_z=0\;\forall z}= 
\left\{
\begin{array}{lc}
\displaystyle  \gamma   & i=y  \\
\displaystyle -\gamma_1 & <iy> \\
\displaystyle -\gamma_2 & <<iy>> \\
0 & \mbox{otherwise}
\end{array}
\right. 
\label{Ginv}
\end{equation}
with 
\begin{displaymath}
\gamma= 1 - \frac{4}{1-l^2} + (1+k_2+k_3)
\frac{1+2c+d}{(1+4l+2c+d)(1-4l+2c+d)} + 
\end{displaymath}
\begin{equation}
+ (1-k_2+k_3)\frac{1}{1-2c+d}+(1-k_3)\frac{2}{1-d}-
\frac{8k_1l}{(1+4l+2c+d)(1-4l+2c+d)}
\label{eq:gamma}
\end{equation}
\begin{displaymath}
\gamma_1=\frac{l}{1-l^2}-\frac{2l}{(1+4l+2c+d)(1-4l+2c+d)}+
\end{displaymath}
\begin{equation}
+k_1\left(\frac{1+2c+d}{(1+4l+2c+d)(1-4l+2c+d)}-\frac{1}{1-d}
\right)-\frac{2(k_2+k_3)l}{(1+4l+2c+d)(1-4l+2c+d)} 
\label{eq:gamma1}
\end{equation}
and
\begin{displaymath}
\gamma_2=\frac{1}{4}(1+k_2+k_3)
\frac{1+2c+d}{(1+4l+2c+d)(1-4l+2c+d)}+
\end{displaymath}
\begin{equation}
-\frac{1}{4}(1-k_2+k_3)\frac{1}{1+2c+d}-\frac{k_2}{2(1-d)}
-\frac{2k_1l}{(1+4l+2c+d)(1-4l+2c+d)} 
\label{eq:gamma2}
\end{equation}
The inverse of the 
Fourier transform of the matrix (\ref{Ginv})
gives the pair correlation function in Fourier space, or structure factor.
It reads as
\begin{equation}
S({\bf p})=\frac{1}{\gamma -\gamma_1\sum_{\mu=1}^2\cos (p_{\mu})
-2\gamma_2(\cos (p_1+p_2)+\cos(p_1-p_2))}
\label{eq:struc-2d}
\end{equation}
\par
\vskip 1cm
{\it 3.3 The disorder line.}
\par
The line where the coefficient of $p^2$ in the denominator of 
(\ref{eq:struc-2d})
is zero is the socalled Lifshitz line where the structure factor
develops a maximum at a value of $p$ different from zero.
It is given by $\gamma_1 + 2 \gamma_2=0$ and is reported in Fig.
\ref{fig:diagr-2d}.
The other interesting line for the behavior of the correlation
function is the disorder line where the correlation
in real space changes its behavior from a purely exponential decay to
an oscillating exponentially decay.
In order to calculate this line it is convenient to take
the spherical average of the
expansion of (\ref{eq:struc-2d}) at small $\vec p$ 
so that the structure factor can be written as \cite{wid}:
\begin{equation}
S({\bf p})=\frac{S(0)}{1+bp^2+cp^4}
\label{eq:struc-form}
\end{equation}
The spherical average can be simply realized by expanding  (\ref{eq:struc-2d})
until the fourth power of $\vec p$ and
taking $p_x=p_y=p/\sqrt{2}$  \cite{wid}. The result is
\begin{equation}
b=\frac{\gamma_1 + 2\gamma_2}{\gamma - 4(\gamma_1+\gamma_2)} \ ;\
c=-\frac{1}{24}\frac{\gamma_1+8\gamma_2}{\gamma - 4(\gamma_1+\gamma_2)}
\label{eq:bbcc}
\end{equation}
On the disorder line the zero of the denominator of 
(\ref{eq:struc-2d}) change from pure imaginary to complex. This happens
when $b^2 - 4c =0$. In Fig. \ref{fig:diagr-2d} we plot the disorder line 
obtained by eqs. (\ref{eq:bbcc}). 

In some bidimensional model with competing interactions
the disorder line coincides with a locus (One-Dimensional-Line) 
where the model can be solved exactly
and has  typical one-dimensional correlations \cite{Ste}.
The  ODL line has been calculated for the model 
(\ref{eq:ham-8vert})  with $J_3=0$
and is given by \cite{Pesc}
\begin{equation}
\cosh 2 J_1 = \frac {e^{4 J_2} + e^{- 4 J_2} +2 e^{-2 J_2}}
{2(1 + e^{2 J_2})}
\label{eq:disli}
\end{equation}
It would be interesting to know whether this line, also reported in Fig. 1, 
coincides with the disorder line of the model 
(\ref{eq:ham-8vert}).
In  the paper \cite{Ent}  
the ODL line has been compared to the disorder line obtained in mean field 
approximation.
Here we have an approximation which correctly reproduce the 
topology of the phase
diagram of the model also at low temperatures. This allows us to compare our
approximation for the disorder line with the expression (\ref{eq:disli})
at any value of $J_1, J_2$. We see that the disorder line calculated by CVM
is very close to the ODL eq. (\ref{eq:disli}), so that it is probably
true that the two lines coincide. For further comparison 
we have plotted in Fig.1 also the disorder line obtained by mean-field 
approximation.

\section{The 3D Ising model with NN, NNN and plaquette interactions}

In this section we study in the CVM approximation
the phase diagram and the disorder line of the model 
(\ref{eq:ham-8vert}) defined on the cubic lattice.
The ground states of the model (\ref{eq:ham-8vert})
are shown in Fig. 2. They can be simply obtained
by minimizing for each value of the parameters 
$J_1,J_2,J_3$ the energy of a single cube
\cite{m1}. 
When two or more cube configurations have the same minimum
energy,
it may be  possible to construct a set of degenerate ground states 
by tiling the whole lattice with the degenerate cube configurations.
We have chosen to study the particular cases $J_2=J_3$ and 
$J_3=-2J_2$ since in these cases the structure of the ground states
is of particular interest for the applications \cite{m1,m2}.
The diagrams of the ground states for these cases are shown in Fig. 3. 

Due to the fact that  the ground states can be expressed in terms of 
single cube configurations, 
it is necessary for a correct implementation of the CVM approximation to 
consider the cube with 8 independent
site magnetizations as 
the maximal  cluster in the CVM implementation.
Then the  approximated free-energy  is given by
\begin{equation}
\begin{array}{ll}
\beta{\cal F} \ &=\ -\; J_1\sum_{<xy>}\, 
{\rm Tr}(\sigma_x\sigma_y\,\rho_{<xy>})\; -\;
J_2\sum_{<<xy>>}\,{\rm Tr}(\sigma_x\sigma_y\,\rho_{<<xy>>})\; + \\
&  -\;J_3\sum_{^x_y\square^w_z}\, {\rm Tr}(\sigma_x\sigma_y\sigma_w\sigma_z
\,\rho_{^x_y\square^w_z})\; 
-\; \beta\sum_x\, h_x{\rm Tr}(\sigma_x\rho_x) \; +\; 
\sum_c\,{\rm Tr} {\cal L}(\rho_c) +\; \\
& -\;
\sum_{^x_y\square^w_z}\,{\rm Tr} {\cal L}(\rho_{^x_y\square^w_z})
+\;\sum_{<xy>}\,{\rm Tr}{\cal L}(\rho_{<xy>})
-\; \sum_x\,{\rm Tr}{\cal L}(\rho_{x})\\
\label{eq:fcubo-acca}
\end{array}
\end{equation}
where $\rho_c$ is the density matrix of an elementary cube.
Also here, as in the bidimensional case, the coefficients of the entropic terms
have been found by applying the prescriptions of ref\cite{An}.

\par
\vskip 1cm
{\it 4.1 The phase diagram with  $J_2=J_3$}

This case has been studied in \cite{m1,m2} by using  mean-field approximation 
and Monte Carlo simulations. In terms of dual surfaces, it corresponds
to the case where the curvature, defined as the number of adjacent plaquettes
forming a  right angle,
 is not weighted \cite{m1}. 
The phase diagram found by applying the natural iteration 
scheme to the  CVM free-energy (\ref{eq:fcubo-acca})
is shown in Fig. \ref{fig:diagr-3d-1}. At positive $J_2$ the ferromagnetic 
phase is stable. On the transition line separating the ferromagnetic and the
paramagnetic phases there is a tricritical point at $J_1^{\it tr}=0.029, 
J_2^{\it tr}=0.0635$.
The coordinates of the tricritical point in mean field approximation and
by Monte Carlo simulations are respectively given by 
$J_1^{{\it tr},MF}=0.11, J_2^{{\it tr},MF}=0.0275$,
$J_1^{{\it tr},MC}=0.03, J_2^{{\it tr},MC}=0.064$. 
A relevant  difference between the mean-field and the CVM phase diagram is
that in the  CVM approximation the paramagnetic phase extends until zero 
temperature, in agreement with results of Monte Carlo simulations.
At negative $J_2$ the  ordered stable configurations
can be constructed starting by the cubes in Fig. \ref{fig:fasitfin}.
The phase between the phases ${\bar 4}$ and $7$ in Fig. 4 has the same 
configuration of phase $7$ but with the magnetizations $m=0$. This phase is 
separated from the phases $\bar 4$ and $7$ by a first-order and a critical 
line, respectively. The phases $\bar 4$ and $7$
are bicontinuous in the sense that the domains 
of spins of one sign form a connected network invading all the lattice
and intertwined with the network formed by the spins of the other sign 
\cite{m1}. 
Also mean field and Monte Carlo simulations show 
stable bicontinuous configuration in this region. 
However, there are discrepancies
between the phase diagrams obtained by different methods at $J_2<0$.
The origin of the discrepancies is probably in the particular  nature of the 
ground states at  $J_2 < -|J_1|/4$. As it can be seen from Fig. 
\ref{fig:circ_gr_st}, the cube configurations $4,\bar 4,7$ are degenerate for 
$J_2 < -|J_1|/4$ and  
an infinite number of bicontinuous 
ground states  can be built up using these configurations.
Here, probably, a low-temperature expansion is needed to understand the 
correct nature of the stable phases, 
but this will be the matter for a future study.
It is interesting to observe that on the transition line between 
the paramagnetic phase and the $\bar 4$  phase at negative $J_2$
there is a tricritical point located at $J_1=0.87, J_2=-0.32$.
It would be interesting to check
the existence of this tricritical point by Monte Carlo simulations.
\par
\vskip 1cm
{\it 4.2  The disorder line.}
\par
In three-dimensions the calculation of the disorder line 
proceeds in the same two steps 
as in  the bidimensional case but with complications due to the higher number
of variational parameters in the $\rho_c$ expansion. This implies
 the existence of many terms like the quantities $k_1$ , $k_2$ , etc. of the 
previous section to evaluate.

The first step is the calculation of the 
elements of the inverse correlation matrix 
(\ref{eq:matrix-acca}), 
that in tree dimensions becomes
\begin{equation}
\beta\left(\frac{\partial h_i}{\partial m_y}\right)_{m_z=0\;\forall z}= 
\left\{
\begin{array}{lc}
\displaystyle  \gamma   & i=y  \\
\displaystyle -\gamma_1 & <iy> \\
\displaystyle -\gamma_2 & <<iy>> \\
\displaystyle -\gamma_3 & <<<iy>>> \\
0 & \mbox{otherwise}
\end{array}
\right. 
\label{Ginv-3D}
\end{equation}
The explicit expressions of the coefficients $\gamma_i$ 
are given in Appendix A for 
clearness purposes.
Going on with the second part of the calculation, we use these matrix elements 
to compute the coefficients (\ref{eq:bbcc}) of the small-${\vec p}$ expansion 
of the structure factor, that in three dimensions are given by:
\begin{equation}
b=\frac{\gamma_1+4\gamma_2+4\gamma_3}
{\gamma-6\gamma_1-12\gamma_2-8\gamma_3}
\; ;\; c=-\frac{1}{36}\frac{\gamma_1+16\gamma_2+28\gamma_3}
{\gamma-6\gamma_1-12\gamma_2-8\gamma_3}
\label{eq:bbcc-3D}
\end{equation}
Also here, as in the bidimensional case, a spherical average has been performed
taking $p_x=p_y=p_z=p/\sqrt{3}$. 
When $b^2-4c<0$ the correlation function in real space is given by
\begin{equation}
G(r)=\frac{{\rm const}}{r}e^{-r/\xi}\sin\frac{2\pi  
r}{\delta}
\label{eq:corr}
\end{equation} 
where
\begin{equation}
\delta=\frac{2\pi}{\left[\frac{1}{2\sqrt{c}}+\frac{b}{4c}\right]^{1/2}}
\;\;\;
\xi=\frac{1}{\left[\frac{1}{2\sqrt{c}}-\frac{b}{4c}\right]^{1/2}}
\;\; .
\label{eq:par-corr}
\end{equation}
The disorder line corresponds to the condition 
$b^2-4c=0$. The results for the disorder line and for the
Lifshitz line $b=0$ are shown in Fig. \ref{fig:diagr-3d-1}.
At low temperatures both the lines behave as $J_2=-J_1/4$.
\par
\vskip 1cm
{\it 4.3  The phase diagram at $J_3=-2 J_2$.}
\par
The ground-state structure corresponding to this case is shown in Fig. 3.
We see that the ferromagnetic phase is bounded from a phase
with spin of different sign in alternate planes. We call this last phase 
a lamellar phase. At finite temperature the phase diagram   
has been  studied in mean field approximation
in a previous paper \cite{m2}. It can well
describe the phase diagram of experimental mixtures of oil, 
water and surfactant.
The CVM approximation is the same used before and the calculation of the
disorder line also proceeds in the way already discussed in the preceding 
paragraph of this section. Therefore we can go on by describing the phase
diagram shown in Fig. \ref{fig:diagr-3d-2}.

At low temperatures and positive $J_2$ the stable phase is the ferromagnetic 
one, while for negative $J_2$ we find the lamellar phase being stable.
On the transition line between the ferromagnetic and the paramagnetic phases
there is a tricritical point at $J_1=0.273$, $J_2=-0.049$, with the
first-order part of the line joining the boundary of the lamellar phase region
at $J_1=0.305$, $J_2=-0.077$ (4-phase point).
This feature, together with the observation that the disorder line intersects
the first-order ferro-paramagnetic transition line in $J_1=0.29$, $J_2=-0.065$,
amidst the tricritical and the 4-phase point, 
 well represents the experimental fact that the microemulsion 
phase (corresponding to the region of the paramagnetic phase below the disorder 
line) can coexist with
the ordered homogeneous phases \cite{m2,GS}. 
We also observe that the Lifshitz line intersects the ordered phases at the 
point $J_1=0.2965\pm .0005$; $J_2=-0.00705\pm .00005$.
 The region close to the 4-phase point is shown in the inset of Fig. 7.

\section{Conclusions.}

The CVM approximation is generally used for studying the topology
of  phase diagrams of spin systems. 
In this paper we have focused on the calculation 
of the correlation functions in the framework of this approximation. 
This calculation can become complicate 
expecially in the case of three-dimensional systems and we have discussed
the origin of these complications. As particular models we have considered
the Ising model with NN, NNN, and plaquette interactions on the square and on 
the cubic lattice. We have calculated the pair correlation function in 
the paramagnetic phase of these models.
We have also given new results about the phase diagrams of these
models for choices of parameters useful for the description of 
experimental systems. 

In particular, in the two-dimensional case with zero plaquette interaction, 
we have calculated the disorder line in the paramagnetic phase and 
we have seen that it is very close to the One-Dimensional Line where
the model is exactly solvable.
This confirms the conjecture that the disorder line and the ODL line
coincide also in this model.
In the three-dimensional case we have chosen two planes in the parameter
space where the phase diagram may have a relevance for the description
of systems of interacting surfaces with fluctuating topology. Fluid mixtures
of oil, water and surfactant are an example of these systems.
We have calculated the disorder line and the Lifshitz line which limit
regions of the paramagnetic phase detectable through scattering experiments.
Our results, shown in Fig. \ref{fig:diagr-3d-1},\ref{fig:diagr-3d-2} confirm 
that the 
three-dimensional Ising model
with NN, NNN, and plaquette interactions can describe many of the 
experimental features appearing in complex fluid mixtures.

\section{Appendix A}

\begin{figure}
\vskip -10mm
\inseps{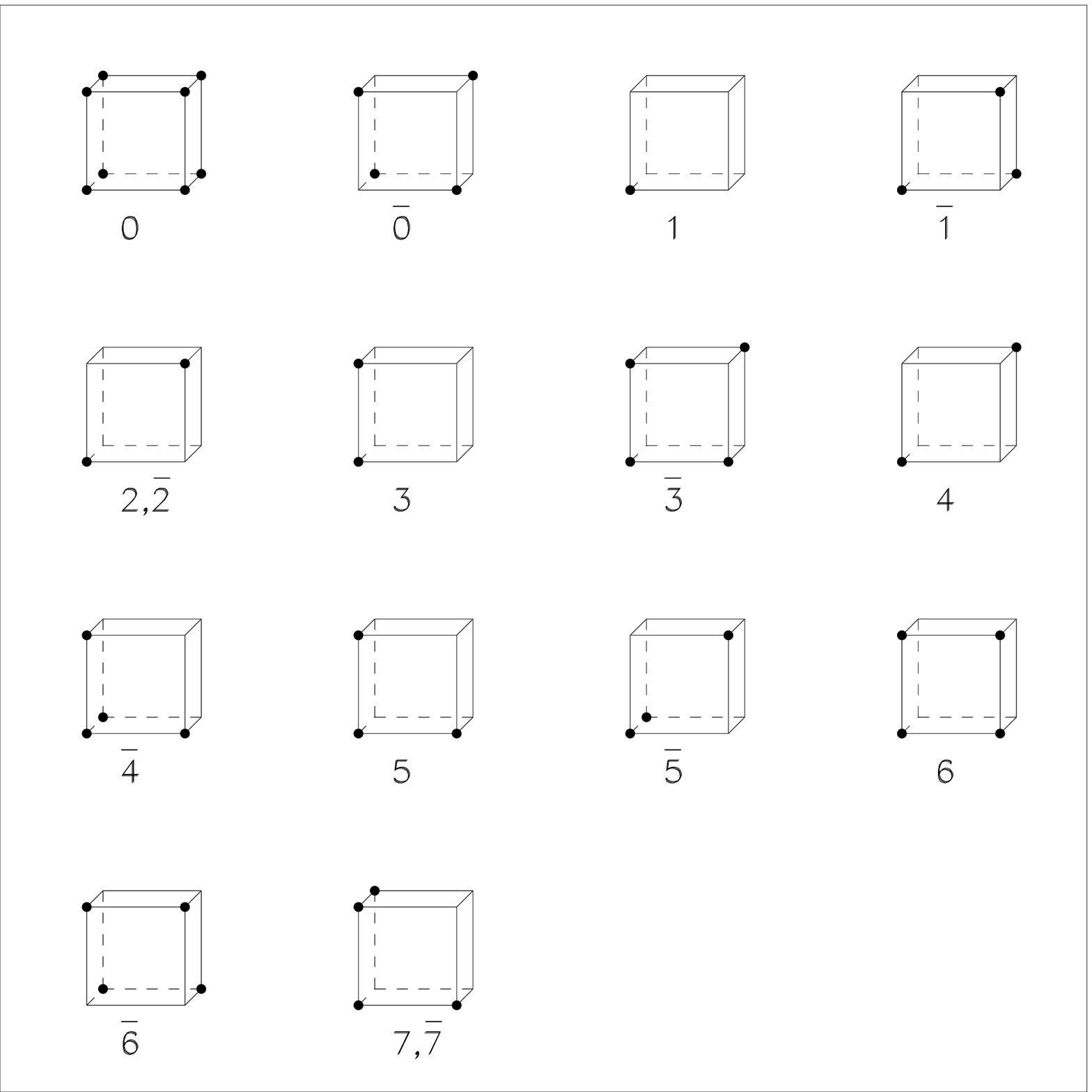}{0.75}
\vskip +20mm
\caption[Testo che va nell'elenco]{
}
\label{fig:gr_states}
\end{figure}

\begin{figure}
\vskip -10mm
\inseps{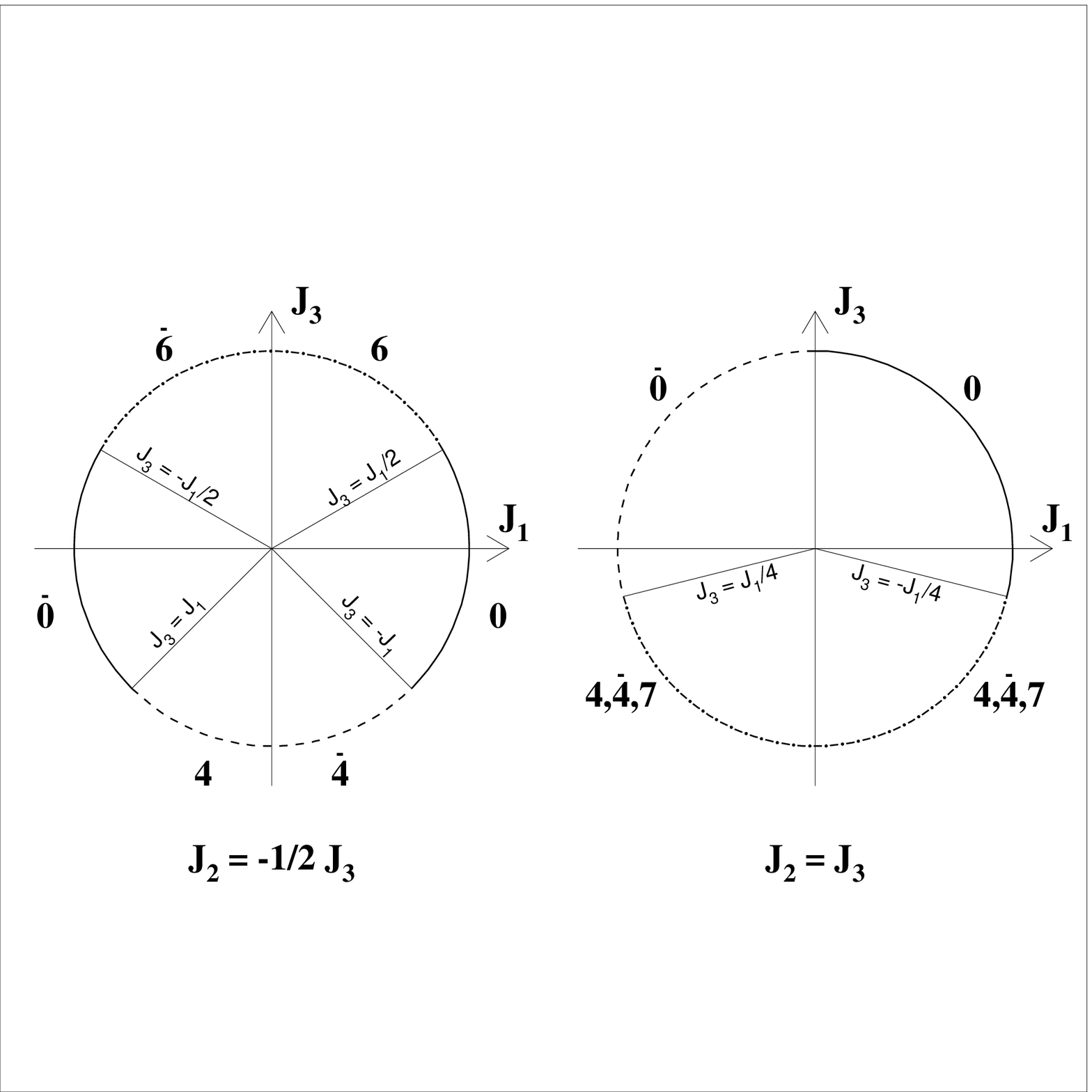}{0.75}
\vskip +20 mm
\caption[Testo che va nell'elenco]{
}
\label{fig:circ_gr_st}
\end{figure}

The matrix (\ref{Ginv-3D}) can be evaluated through the same procedure 
shown in sect. III, starting from the single spin magnetization state equation 
(\ref{eq:acca_i}), that in 3D turns out to be:
\begin{eqnarray}
\beta h_i
&=& \frac{1}{256}\sum_{c\ni i}{\rm Tr}(\sigma_i\log\rho_c)
-\frac{1}{16}\sum_{^i_y\square^w_z}{\rm Tr}(\sigma_i
              \log\rho_{^i_y\square^w_z}) \nonumber\\
& & +\frac{1}{4}\sum_{<iy>}{\rm Tr}(\sigma_i\log\rho_{<iy>})
-\frac{1}{2}{\rm Tr}(\sigma_i\log\rho_i)
\label{eq:acca_i-3D}
\end{eqnarray}
Differentiating this equation with respect to the site magnetizations,
we realize that we have to deal with a long list of derivatives of the
kind: $D_i^{\alpha}=\frac{\partial\xi_{\alpha}}{\partial m_i}$ where \
$\xi_{\alpha}={\rm Tr}(\sigma_{\alpha}\rho_{\alpha})$ is the parameter in the 
expansion of the density matrix corresponding to the subcluster $\alpha$.
By symmetry considerations and taking into account that in the paramagnetic 
phase the only derivatives different from zero are those from  
subclusters with an odd number of sites, we can reduce the 
number of derivatives to 33, listed  in Fig. \ref{fig:derivate}. 
Repeating the procedure used in section III to obtain (\ref{eq:system-kappa}), 
we can get to a system of 33 linear equations which, although very long, can be 
straightforwardly solved and it will not be reported here. 
The solution of this system gives us the coefficients $k_i$ , $a_i$ , $b_i$ 
appearing in Fig. \ref{fig:derivate} and allows us to calculate the 
expression of the matrix elements (\ref{Ginv-3D}), that 
we write down here explicitly:
\begin{eqnarray}
\gamma\ &=&\ 8\,[\, C^{0c} + 6k_1C^3 + 3k_2C^2 + 3k_3C^6 + 3k_4C^2 + 3k_5C^4 +
3k_6C^3 + 3k_7C^{\bar 6} + \nonumber\\
& & + 6k_8C^7 + 6k_9C^{\bar 3} + k_{10}C^{\bar 4} + 3k_{11}C^2 + 
3k_{12}C^{\bar 3} + k_{13}C^{\bar 0} + 6a_1C^{3c} + 3a_2C^{2c} + \nonumber\\
& & + 3a_3C^6 + 6a_{14}C^7 + 3a_{15}C^{\bar 4} + 
3a_{16}C^{\bar 3} + 3a_4C^{2c} + 3a_5C^{4c} + 3a_6C^{3c} + 3a_7C^{\bar 6} + 
\nonumber\\
& & + 6a_8C^7 + 6a_9C^{\bar 3} + a_{10}C^{3c} +3a_{11}C^{2c} 
+3a_{12}C^{\bar 3} + a_{13}C^{\bar 0} + b_1C^0 + 3b_2C^{3c} + \nonumber\\
& & + 3b_3C^{2c} + b_4C^{4c}\,]\; -\; 12\,[\, P^{0c} + 
2k_1P^1 + k_2P^2 + k_3P^0\,]\; +\; 6L^{0c} \; -\; 1 
\end{eqnarray}
\begin{eqnarray}
-\gamma_1\ &=&\ 4\,[\, C^3 + 2k_1(C^2+C^6+C^7) + k_2(2C^3+C^{\bar 4}) + 
k_3(2C^3+
C^{\bar 3}) + k_4(C^4+2C^{\bar 3})   
+ \nonumber\\
& & + k_5(C^2+2C^7) + k_6(2C^7+C^{\bar 6}) + k_7(C^3+2C^{\bar 3}) + 
2k_8(C^3+C^4+C^{\bar 3}) + 2k_9(C^2+ \nonumber\\
& & +C^7+C^{\bar 6}) + k_{10}C^2 + k_{11}(C^{\bar 4}+C^{\bar 3}) 
+ k_{12}(2C^2+C^{\bar 0}) + k_{13}C^{\bar 3} + 2a_1(C^{2c}+C^6+C^7) + 
\nonumber\\
& & + a_2(2C^{3c}+C^{\bar 4}) + a_3(2C^{3c}+
C^{\bar 3}) + 2a_{14}(C^{3c}+C^{\bar 4}+C^{\bar 3}) + a_{15}(C^{2c}+2C^7) + 
\nonumber\\
& & + a_{16}(C^6+2C^7) + a_4(C^4+2C^{\bar 3}) + a_5(C^2+2C^7) 
+ a_6(2C^7+C^{\bar 6}) + a_7(C^3+2C^{\bar 3}) + \nonumber \\
& & + 2a_8(C^3+C^4+C^{\bar 3}) + 2a_9(C^2+C^7+
C^{\bar 6}) + a_{10}C^2 + a_{11}(C^{\bar 4}+C^{\bar 3}) +\nonumber\\
& & + a_{12}(2C^2+C^{\bar 0}) + 
a_{13}C^{\bar 3} + b_1C^{3c} + b_2(C^0+2C^{2c}) + b_3(2C^{3c}+C^{4c}) + 
b_4C^{2c}\, ]\; + \nonumber\\
& & -\; 4\,[\, P^1 + k_1(P^2+P^{0}) + k_2P^1 + k_3P^1\, ]\; +\; L^0
\end{eqnarray}
\begin{eqnarray}
-\gamma_2 &=& 2\,[\, C^2 + 2k_1(C^3+C^{\bar 4}+C^{\bar 3}) + k_2(C^6+2C^7) + 
k_3(C^2+2C^7) + k_4(2C^7+C^{\bar 6}) + \nonumber\\
& & + k_5(C^3+2C^{\bar 3}) + k_6(C^4+2C^{\bar 3}) + k_7(C^2+2C^7) + 
2k_8(C^2+C^7+C^{\bar 6}) + \nonumber\\
& & + 2k_9(C^3+C^4+C^{\bar 3}) + k_{10}C^{\bar 3} + k_{11}(2C^2+C^{\bar 0}) +
k_{12}(C^{\bar 4}+2C^{\bar 3}) + k_{13}C^2 + \nonumber\\
& & 2a_1(C^{3c}+C^{\bar 4}+C^{\bar 3}) + a_2(C^6+2C^7) + a_3(C^{2c}+2C^7) + 
2a_{14}(C^{2c}+C^6+C^7) + \nonumber\\
& & + a_{15}(2C^{3c}+C^{\bar 3}) + a_{16}(2C^{3c}+C^{\bar 4}) + 
a_4(2C^7+C^{\bar 6}) 
+ a_5(C^{3c}+2C^{\bar 3}) + \nonumber\\
& & + a_6(C^{4c}+2C^{\bar 3}) + a_7(C^{2c}+2C^7) + 
2a_8(C^{2c}+C^7+C^{\bar 6}) + 2a_9(C^{3c}+C^{4c}+C^{\bar 3}) 
+ \nonumber\\
& & a_{10}C^{\bar 3} + a_{11}(2C^{2c}+C^{\bar 0}) + a_{12}(C^{\bar 4}+
2C^{\bar 3}) + a_{13}C^{2c} + b_1C^{2c} + b_2(C^{4c}+2C^{3c}) + \nonumber\\
& & + b_3(C^0+2C^{3c}) + b_4C^{3c} \,]\; -\; [P^2 + 2k_1P^1 + k_2P^0 + 
k_3P^2] 
\end{eqnarray}
\begin{eqnarray}
-\gamma_3 &=& C^4 + 6k_1C^7 + 3k_2C^{\bar 3} + 3k_3C^{\bar 4} + 3k_4C^{0c} +
3k_5C^{\bar 6} + 3k_6C^2 + 3k_7C^4 + 6k_8C^{\bar 3} + \nonumber\\
& & + 6k_9C^{\bar 3} + k_{10}C^{\bar 0} + 3k_{11}C^{\bar 3} +
+ 3k_{12}C^2 + k_{13}C^{\bar 4} + 6a_1C^7 + 3a_2C^{\bar 3} + 3a_3C^{\bar 4} 
+ \nonumber\\
& & + 6a_{14}C^{3c} + 3a_{15}C^6 + 3a_{16}C^{2c} + 3a_4C^{0c} + 
3a_5C^{\bar 6} + 3a_6C^{2c} + 3a_7C^{4c} + 6a_8C^{\bar 3} 
+ \nonumber\\
& & + 6a_9C^{\bar 3} + a_{10}C^{\bar 0} + 3a_{11}C^{\bar 3} + 3a_{12}C^{2c} + 
a_{13}C^{\bar 4} + b_1C^{4c} + 3b_2C^{2c} + 3b_3C^{3c} + b_4C^0 \nonumber\\
\end{eqnarray}
where we used the notations:
$
P^0 = \frac{1}{256}\;{\rm Tr}\left(\frac{\sigma_x\sigma_y\sigma_z\sigma_w}
{\rho_{^x_y\square^w_z}}\right)$, 
$
P^1 = \frac{1}{256}\;{\rm Tr}\left(\frac{\sigma_i\sigma_j}
{\rho_{^i_j\square^w_z}}\right)$, 
$
P^2 = \frac{1}{256}\;{\rm Tr}\left(\frac{\sigma_i\sigma_j}
{\rho_{^i_y\square^w_j}}\right)$,
$
P^{0c} = \frac{1}{256}\;{\rm Tr}\left(\frac{1}
{\rho_{^x_y\square^w_z}}\right)$, 
$
L^0 = \frac{1}{16}\;{\rm Tr} 
\left(\frac{\sigma_x\sigma_y}{\rho_{<xy>}}\right)$,
$
L^{0c} = \frac{1}{16}\;{\rm Tr}\left(\frac{1}{\rho_{<xy>}}\right)$ 
and 
$
C^{\alpha} = \frac{1}{(256)^2}\;{\rm Tr}\left(\frac{\sigma_{\alpha}}{\rho_c}
\right)$, 
$\alpha$ being the name of the cluster as they are classified in 
Fig. \ref{fig:gr_states}, and 
$\sigma_{\alpha} = \prod_{i\in\alpha} \sigma_i$, as in Section II.


\newpage
\newpage
\newpage
\newpage
\begin{figure}
\vskip -10mm
\inseps{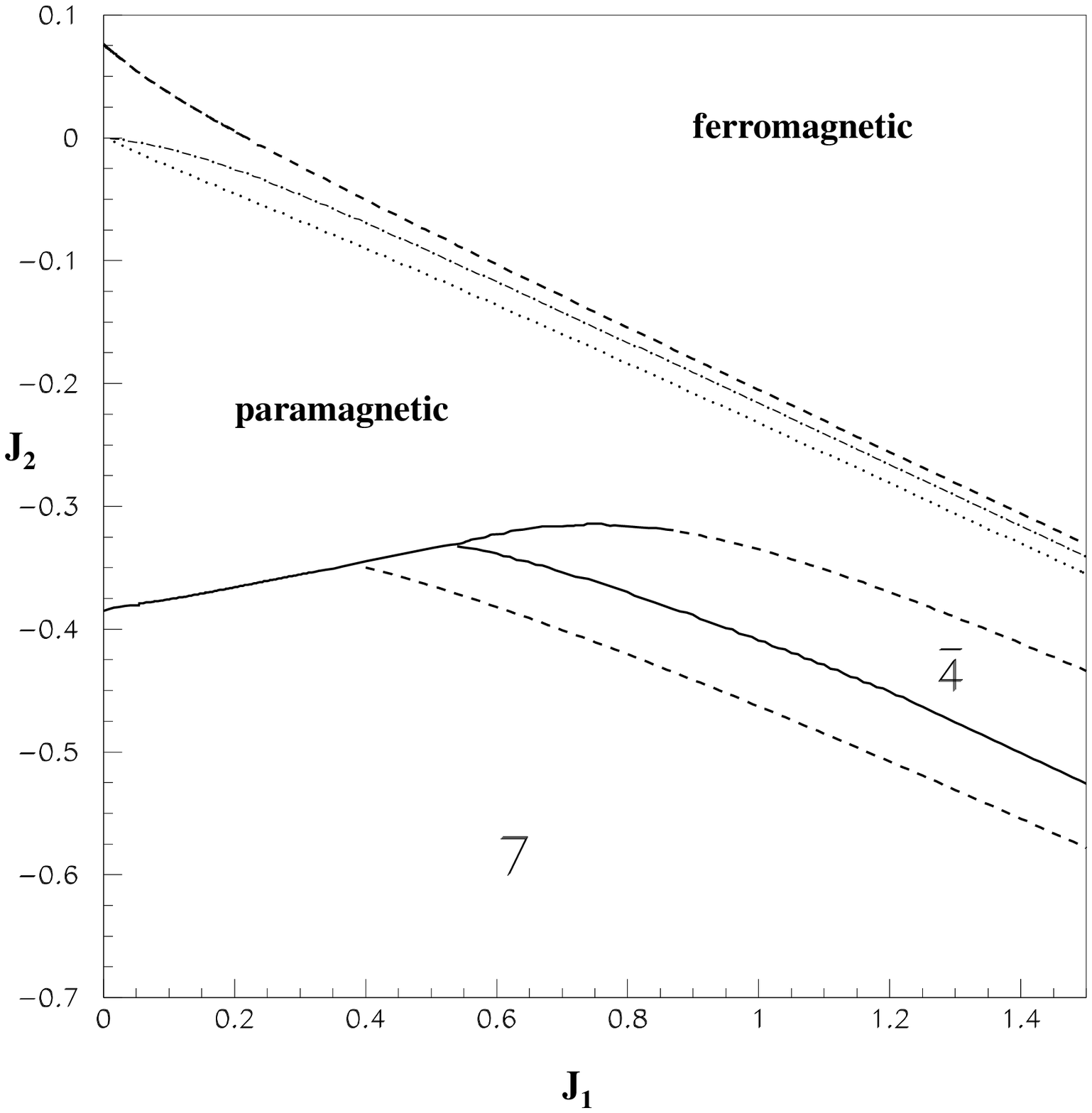}{0.75}
\vskip +20mm
\caption[Testo che va nell'elenco]{
}
\label{fig:diagr-3d-1}
\end{figure}

\newpage
\begin{figure}
\vskip -10mm
\inseps{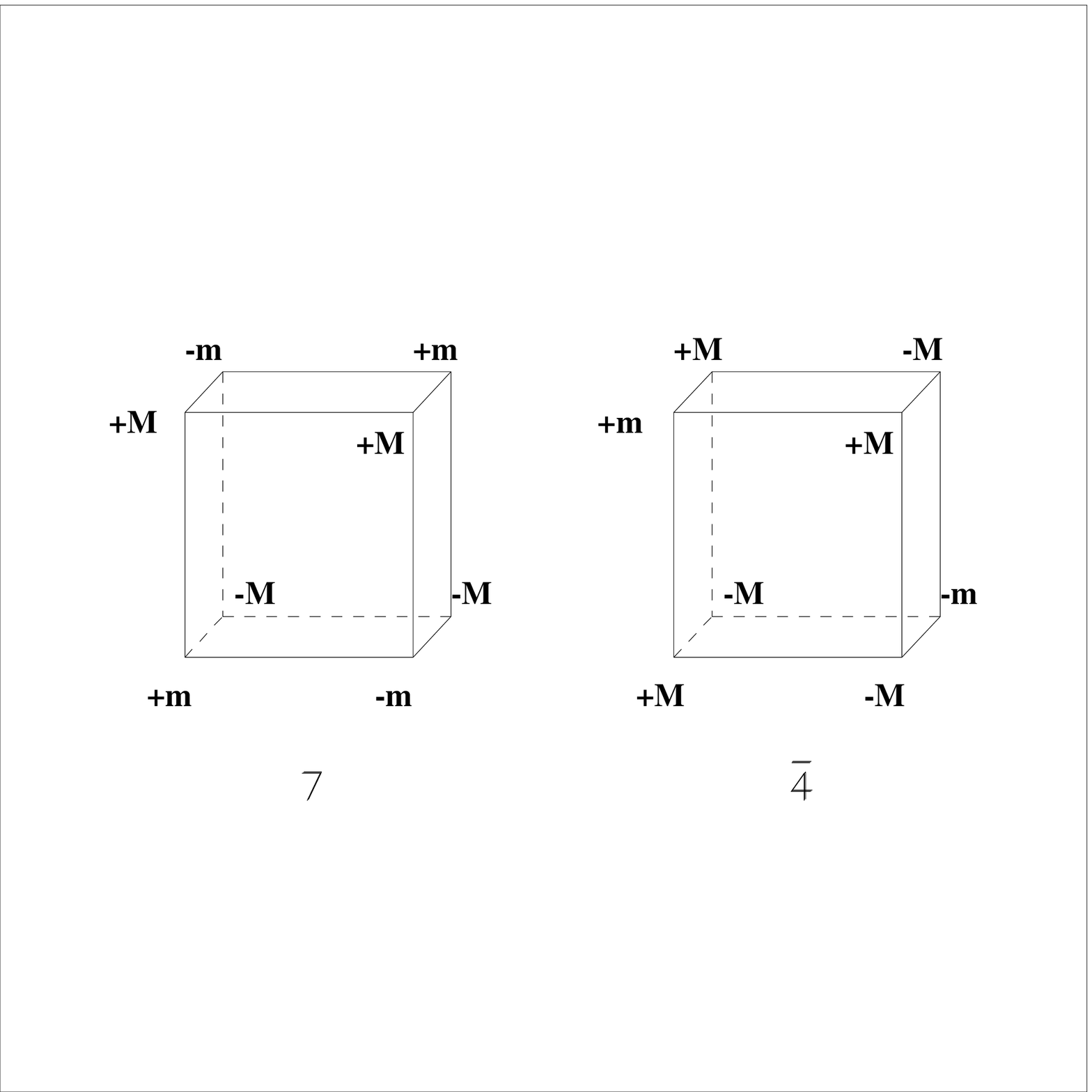}{0.75}
\vskip +20mm
\caption[Testo che va nell'elenco]{
}
\label{fig:fasitfin}
\end{figure}
\newpage
\begin{figure}
\vskip -10mm
\inseps{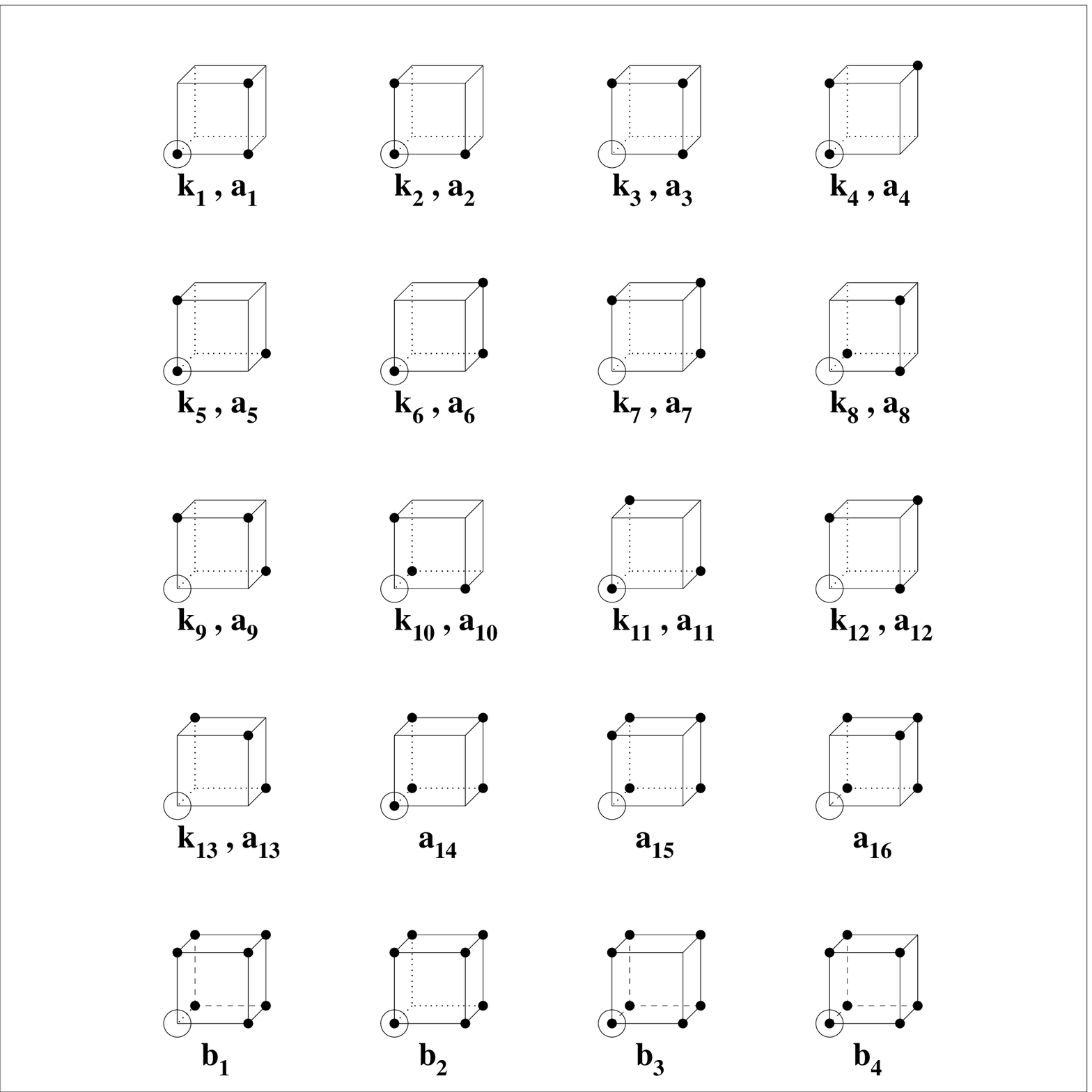}{0.75}
\vskip +20mm
\caption[Testo che va nell'elenco]{
}
\label{fig:derivate}
\end{figure}
\newpage
\begin{figure}
\vskip -10mm
\inseps{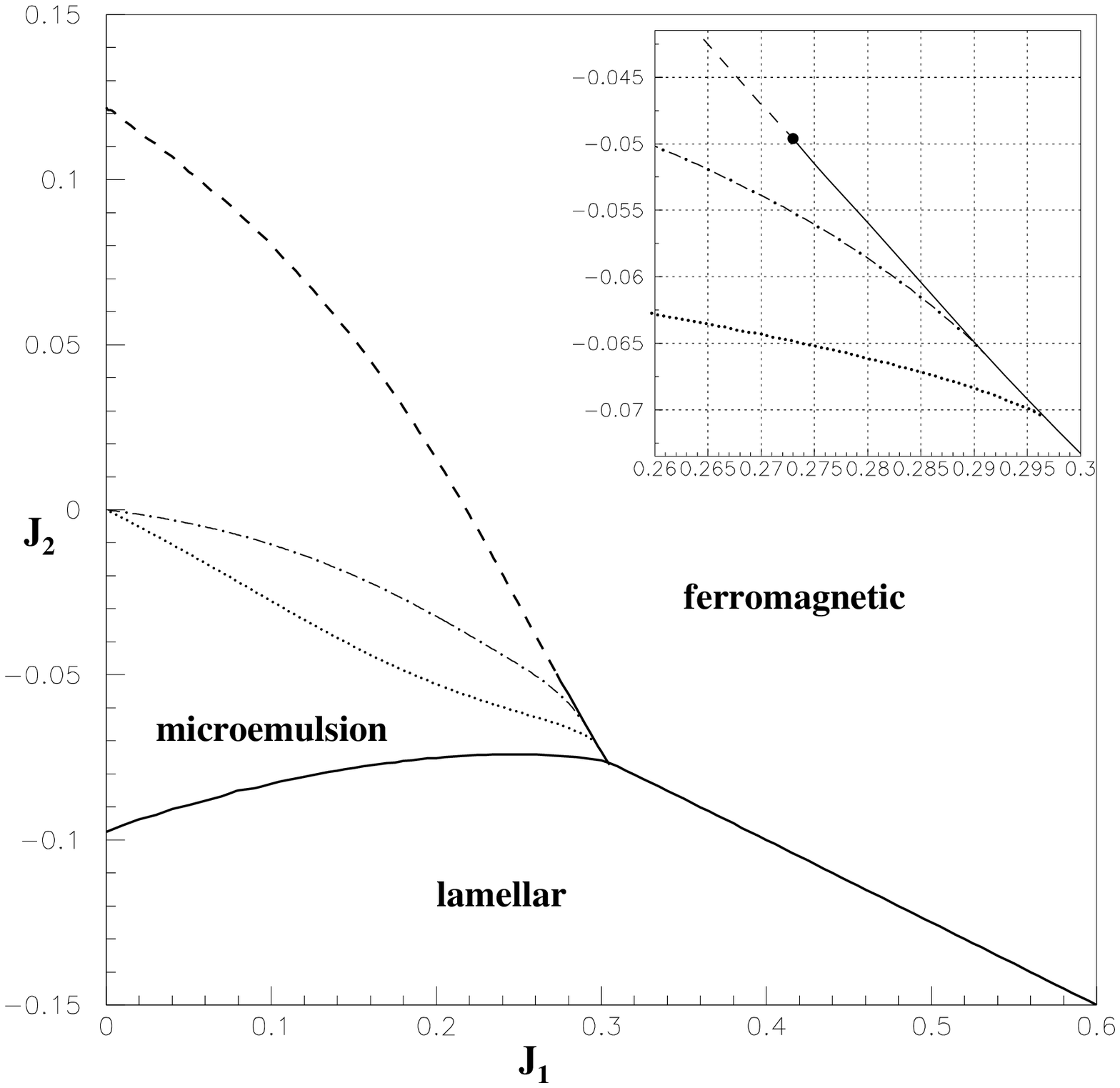}{0.75}
\vskip +20mm
\caption[Testo che va nell'elenco]{
}
\label{fig:diagr-3d-2}
\end{figure}

\newpage
\centerline{\bf Figure captions}
\vskip 1 truecm
\noindent{\bf Fig.1:}
Phase diagram of model (\ref{eq:ham-8vert}) in $D=2$ with $J_3=0$. The dashed
lines
are critical lines separating the disordered phase from the ferromagnetic
and the lamellar (SAF) phases. The first-order transition between the
paramagnetic and SAF phases is represented by a solid line.
The dotted line is the one-dimensional line of equation (\ref{eq:disli}),
while the light dashed and dash-dotted lines are the mean field and plaquette
approximation
calculation of the disorder line, respectively.

\vskip 0.5 truecm
\noindent{\bf Fig.2:}
Ground states of the 3D model. The dot $\bullet$ indicates $\sigma=+1$
while $\sigma=-1$ in the remaining sites. The labels refer both to the phase
and to the dotted subclusters. The complementary (non-dotted) subclusters can
be referred to adding a ``c" after the label.

\vskip 0.5 truecm
\noindent{\bf Fig.3:}
Ground states of the 3D model for the values $J_2=J_3$ and $J_2=-J_3/2$ of the
coupling constants.

\vskip 0.5 truecm
\noindent{\bf Fig.4:}
Phase diagram of model (\ref{eq:ham-8vert}) in $D=3$ with $J_3=J_2$.
The dash-dotted line is the disorder line in cube approximation.
The dotted line is the Lifshitz line. In the narrow strip between the $7$ and
$\bar 4$ phases we find a type-7 phase with the magnetizations $m$ of Fig. 5
equal to zero. As before, solid and dashed lines represent first-order and
critical transition lines.

\vskip 0.5 truecm
\noindent{\bf Fig.5:}
Phases at finite temperature appearing in the diagram of Fig. 
\ref{fig:diagr-3d-1}.
Capital letters indicate bigger magnetizations.

\vskip 0.5 truecm
\noindent{\bf Fig.6:}
Derivatives $D_i^{\alpha}=\frac{\partial\xi_{\alpha}}{\partial m_i}$
with respect to the magnetization $m_i$ in the site circled (see App. A
for the notations). $k_i$, $a_i$ and $b_i$ are respectively the derivatives of
the 3-, 5- and 7-sites subclusters, represented by the dots.
>From $k_1(a_1)$ to $k_3(a_3)$ the cluster considered is $\alpha=5(5c)$;
from $k_4(a_4)$ to $k_9(a_9)$, $\alpha={\bar 5}({\bar 5}c)$, from
$k_{10}(a_{10})$ to $k_{13}(a_{13})$, $\alpha={\bar 1}({\bar 1}c)$,
from $a_{14}$ to $a_{16}$, $\alpha=5c$ and there is no correspondent in the
$k_i$'s; from $b_1$ to $b_4$, $\alpha=1c$.

\vskip 0.5 truecm
\noindent{\bf Fig.7:}
Phase diagram of model (\ref{eq:ham-8vert}) in $D=3$ with $J_3=-2J_2$.
The dash-dotted line is the disorder line; the dotted line is the Lifshitz line.
In the inset, the region where the disorder and the Lifshitz lines cross the
first-order Ferro-Paramagnetic transition line is enlarged.

\end{document}